\documentclass[]{tEWA2e}
\usepackage{color}
\usepackage[mathscr]{euscript}

\begin{document}

\doi{10.1080/0950034YYxxxxxxxx}

\markboth{V. I. Fesenko, et al.}{Journal of Electromagnetic Waves and Applications}


\title{Control of single-mode operation in a circular waveguide filled by a~longitudinally magnetized gyroelectromagnetic medium}

\author{Volodymyr I. Fesenko$^{a,b,\ast}$\thanks{$^\ast$Corresponding author. Email: volodymyr.i.fesenko@gmail.com \vspace{6pt}}, Vladimir R. Tuz$^{a,b,c}$, Illia V. Fedorin$^{d}$, Hong-Bo Sun$^{c}$, \\ Valeriy M. Shulga$^{a,b}$, and Wei Han$^{a,e}$ \\\vspace{6pt}
$^{a}${\em{International Center of Future Science, Jilin University, Changchun, People's Republic of China}}; $^{b}${\em{Institute of Radio Astronomy of National Academy of Sciences of Ukraine, Kharkiv, Ukraine}}; $^{c}${\em{State Key Laboratory on Integrated Optoelectronics, College of Electronic Science and Engineering, Jilin University, Changchun, People's Republic of China}}; $^{d}${\em{National Technical University `Kharkiv Polytechnical Institute', Kharkiv, Ukraine}}; $^{e}${\em{College of Physics, Jilin University, Changchun, People's Republic of China}}  \\\vspace{6pt}
\received{March 2017} }

\maketitle

\begin{abstract}
A substantial control of dispersion features of the hybrid EH$_{01}$ and HE$_{11}$ modes of a circular waveguide which is completely filled by a longitudinally magnetized composite finely-stratified ferrite-semiconductor structure is discussed. A relation between the resonant conditions of such a composite gyroelectromagnetic filling of the circular waveguide and dispersion features of the supported modes are studied. Three distinct frequency bands with the single-mode operation under normal as well as anomalous dispersion conditions of the EH$_{01}$ mode are identified by solving an optimization problem with respect to the filling factors of the composite medium. The possibility of achieving isolated propagation of the HE$_{11}$ mode is revealed.
\bigskip

\bigskip
\vspace{12pt}

\end{abstract}

\section{Introduction}

Today the general theory of metallic waveguides designed to operate in the microwave part of spectrum is well developed and described in a number of subsequent papers and textbooks \cite{Fuller_book_1969, Collin_book_1991, Pozar_book_2012}. This theory implies a solution of wave equations being under constraints of  boundary conditions related to perfectly conducting (PEC) walls with accounting the guide geometry. The required solution appears as a set of waveguide (eigen)modes which differ by their cutoff frequencies and dispersion. Nevertheless, while the corresponding solutions obtained for hollow and isotropically filled waveguides with coordinate boundaries (e.g. rectangular or circular waveguides) are well known and can be derived in an explicit analytical form, those of the waveguides with an inhomogeneous and/or anisotropic (gyrotropic) filling are much more complicated and require some particular considerations accompanied by cumbersome numerical calculations for each concrete design \cite{kales_JApplPhys_1953, Waldron_RadioEng_1959, Novotny_PhysRevE_1994, Che_JEMWA_2002, Zaginaylov_PlasmaPhysRep_2005, Fesenko_2005, Cojocaru_JOptSocAmB_2010, Tuz_PIERB_2011, Pollock_AppPhys_2016}. Moreover, in the latter case the dispersion properties of such waveguides become to be sufficiently intricate and effects of anomalous dispersion, complex wave propagation, mode coupling, splitting and conversion may occur. On the other hand, such a diversity in dispersion features of the waveguide modes gives an ability to design systems with demanded characteristics. That is the reason why in recent years  main research efforts are directed towards the study of electromagnetic features of waveguides filled by artificially created materials (metamaterials) in order to realize waveguide systems having unique operation conditions including those suitable for transmission in the range of terahertz waves \cite{Brand_IntJElectron_2009, Ghosh_Electromagn_2012, Pollock_MTT_2013}. In particular, in the present paper, extraordinary dispersion features of modes of a circular waveguide filled by a composite medium possessing combined gyroelectromagnetic effect \cite{Wenyan_MTT_1994, prati_JEMWA_2003, Tuz_JEMWA_2017} are of a special interest.

Indeed, conventional hollow and isotropically filled circular waveguides support transverse electric (TE) and transverse magnetic (TM) modes from which TE$_{11}$ mode is the fundamental one, whereas all waves supported by  longitudinally magnetized gyrotropically (ferrite or plasma) filled waveguides are not pure TE and TM modes, since they include longitudinal components of both the electric and magnetic fields. It is convenient to distinguish these waves as hybrid HE and EH modes, depending whether the longitudinal component of either magnetic or electric field is dominant, and, in general, these waves are unable to degenerate into TE and TM modes under the axial symmetry of the guide \cite{fuller_book_1987}. In such a gyrotropic waveguide  HE$_{11}$ mode is the fundamental one which is inherited from the TE$_{11}$ mode of the hollow circular waveguide. Moreover, in such a gyrotropic waveguide, all non-symmetric HE$_{nm}$ and EH$_{nm}$ modes with nonzero index $n$ split into waves acquiring left-handed and right-handed circular polarizations \cite{Gurevich_book_1963, fuller_book_1987}.

It is a standard engineering practice to choose parameters of a circular waveguide in such a way to ensure propagation of only TE$_{11}$ and TM$_{01}$ modes, all other higher order modes are cut off and are non-propagating ones. Thus, operation of the microwave links in the single-mode conditions with supporting only the fundamental mode is particularly useful in tasks of the power transmission over long distances and for broadcasting systems \cite{Wittaker_book_1999}. Besides, nonreciprocal behaviors of the hybrid HE$_{11}$ mode which propagates through a circular waveguide loaded by a longitudinally magnetized ferrite rod are utilized in a wide class of ferrite-based devices such as circulators, rotators, polarized absorbers and duplexers \cite{fuller_book_1987, Pozar_book_2012}. Nevertheless, some specific applications require for a waveguide to operate at the modes other than the fundamental one. For instance, the higher order TE$_{0m}$ modes (especially, the TE$_{01}$ mode) are very attractive for their use for the low-loss power transmission over long distances and in resonant cavities with very high quality factor since these modes possess significantly smaller attenuation with frequency increasing compared to the fundamental mode. The systems in the particle-beam physics operate on the hybrid EH$_{01}$ mode in order to accelerate particles to relativistic speeds in the plasma filled circular waveguides \cite{Kumar_JApplPhys_2008, Cook_PhysRevLett_2009, Pollock_AppPhys_2016}. Interest in the circular waveguides operating on EH$_{01}$/TM$_{01}$ mode is also conditioned by the fact that some of high-power microwave sources (such as the relativistic backward-wave oscillators and the magnetically insulating transmission-line oscillators) generate exactly the TM$_{01}$ circular waveguide mode. 

In spite of prospects of using higher order modes instead of the fundamental one for specific applications, there are several problems observed during their excitation and retention. One of the problems is that these modes are not dominant in circular waveguides. Thus, in order to provide their support, a waveguide should be oversized that inevitably results in appearance of the undesirable propagation of a number of other modes having lower cutoff frequencies. Another issue is that for the higher order modes the coupling between the desired and undesired modes usually appears at the irregularities of the guide (e.g. bands, non-ideal inner cross-section, etc.). Moreover, the simultaneous presence of several modes in the waveguide causes not only higher power losses but also distortion of the transmitted signal due to the inter-modal dispersion effect.

Therefore, there is an interest in designs of waveguides which are able to support only a single desired higher order mode with avoiding its significant distortion and attenuation during  transmission. A possibility of such a single-mode operation for hybrid EH$_{01}$ and HE$_{11}$ modes in a hollow circular waveguide has been recently reported in \cite{Pollock_MTT_2013, Pollock_AppPhys_2016}. In particular, it has been demonstrated that a single-mode operation conditions can be achieved  for the hybrid EH$_{01}$ mode that possesses an anomalous dispersion in the circular waveguide  with either metamaterial-lined PEC wall or metamaterial-coated PEC rod. It is shown that insertion of a thin metamaterial liner which exhibits both dispersive epsilon-negative and near-zero properties into a circular waveguide leads to formation of the passband with anomalous dispersion for the hybrid HE$_{11}$ mode which arises far below its conventional cutoff frequency.

In our previous study \cite{Tuz_JEMWA_2017} it is revealed that in a circular waveguide completely filled by a longitudinally magnetized composite finely-stratified ferrite-semiconductor structure the combined geometrical and material dispersion appears in the hybrid modes which differs drastically from that of conventional dielectric, ferrite or plasma filled waveguides. Moreover, simultaneous presence of both gyromagnetic and gyroelectric effects in the waveguide system allows to gain a substantial control over the dispersion characteristics and field distribution of the supported modes by utilizing an external static magnetic field as a driving agent. As a further elaboration of the results obtained in \cite{Tuz_JEMWA_2017}, in this paper we demonstrate that bands of the single-mode operation for the hybrid EH$_{01}$ and HE$_{11}$ modes can be controlled by a concisions choice of material and geometrical parameters of such a composite gyroelectromagnetic filling of a circular waveguide. 

\section{Outline of problem}
\label{sec:problem}

\begin{figure}[htbp]
\centering
\includegraphics[width=1.0\linewidth]{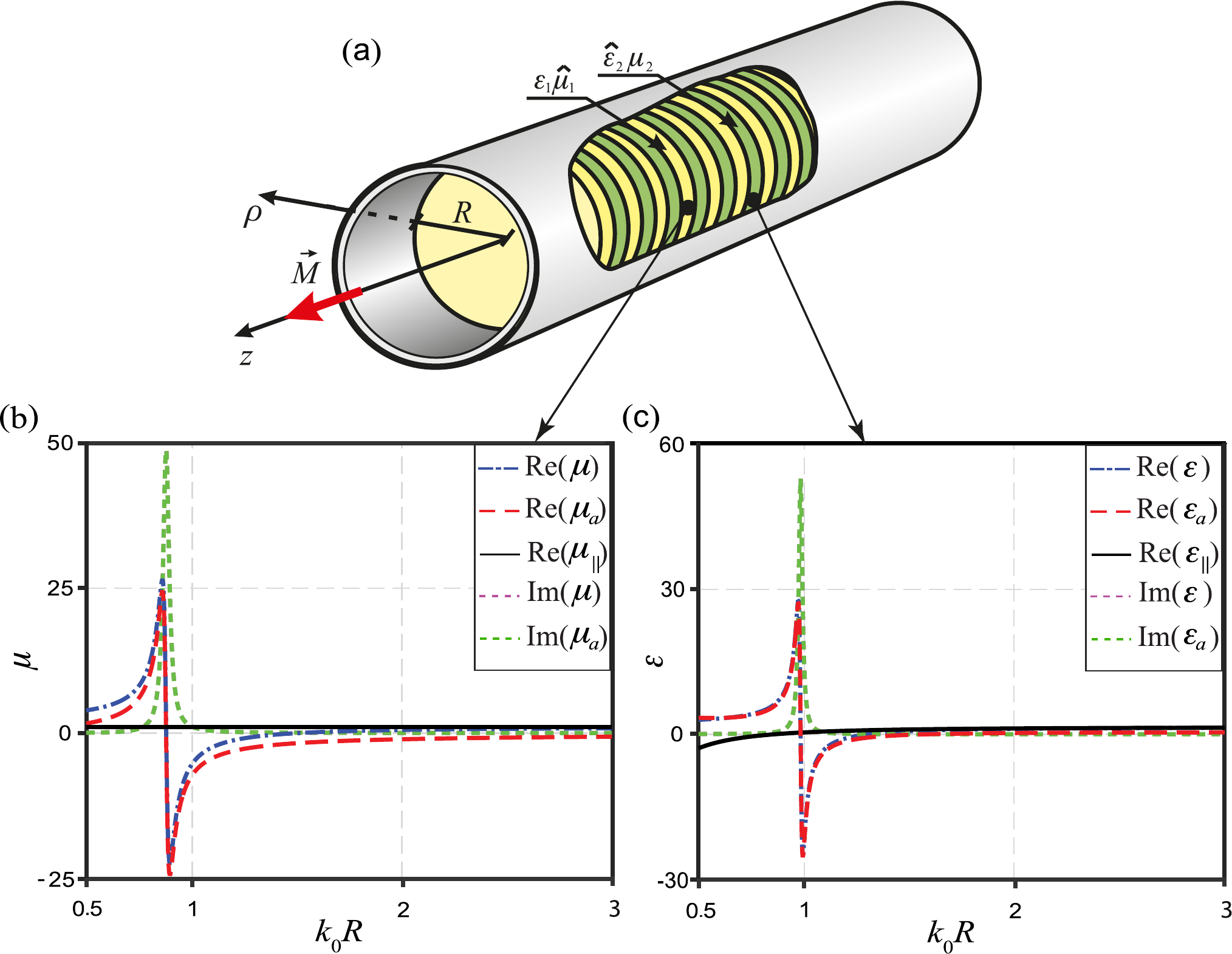}
\caption{(a) Schematic of a circular waveguide completely filled by a longitudinally magnetized multilayered ferrite-semiconductor structure and dispersion curves of the tensors components of (b) ferrite permeability $\hat\mu$ and (c) semiconductor permittivity $\hat\varepsilon$ depicted on the frequency scale normalized on the radius $R$ of the guide; for ferrite, under saturation magnetization of 2930~G, parameters are: $f_0=\omega_0/2\pi=4.2$~GHz, $f_m=\omega_m/2\pi=8.2$~GHz, $\varepsilon_f=5.5$, $b=0.02$; for semiconductor, parameters are: $f_p=\omega_p/2\pi=4.9$~GHz, $f_c=\omega_c/2\pi=4.7$~GHz, $\varepsilon_l=1.0$, $\nu/2\pi=0.05$~GHz, $\mu_s=1.0$.}
\label{fig:fig_1}
\end{figure}

Thereby, in this paper we study dispersion peculiarities of modes of an axial waveguide encircled by a PEC wall with radius $R$ (Figure~\ref{fig:fig_1}a). The waveguide inner region is considered to be completely filled by a longitudinally magnetized composite medium which is constructed by combining together of gyromagnetic (ferrite with constitutive parameters $\varepsilon_f$, $\hat \mu_f$) and gyroelectric (semiconductor with constitutive parameters $\hat \varepsilon_s$, $\mu_s$) layers providing that these constitutive layers are arranged periodically along the guide, i.e., along the $z$-axis. Therefore, the alternating layers form a gyroelectromagnetic superlattice which period is $L$ and thicknesses of the constitutive ferrite and semiconductor layers are $d_f$ and $d_s$, respectively. The filling factors balance $\delta_f + \delta_s = 1$ is additionally introduced in order to define the ratio between the ferrite $\delta_f=d_f/L$ and semiconductor $\delta_s=d_s/L$ fractions of the superlattice. For further reference the dispersion characteristics of the tensor components of permeability $\hat \mu_f$ of the ferrite layers and permittivity $\hat \varepsilon_s$ of the semiconductor layers are presented in panels (b) and (c) of Figure~\ref{fig:fig_1}, respectively. They are calculated  using common expressions for description of constitutive parameters of normally biased ferrite \cite{Gurevich_book_1963, Collin_book_1992} and semiconductor \cite{Bass_book_1997} materials taking into account the losses. In particular, it is supposed that the practical structure is made in the form of a barium-cobalt/doped-silicon superlattice \cite{Wu_JPhysCondensMatter_2007}.

In what follows we stipulate two restrictions. First, the strength of an external static magnetic field $\vec M$ which is aligned along the axis of the guide (i.e., along the $z$-axis) is high enough to produce a homogeneous saturated state of magnetic as well as semiconductor layers. Second, the whole composite filling of the guide is considered to be a finely-stratified structure. It implies that all characteristic dimensions of such a multilayered ferrite-semiconductor structure are much smaller than the wavelength in the corresponding parts of the composite medium, i.e., $d_f\ll \lambda$, $d_s \ll \lambda$, and $L \ll \lambda$ (the long-wavelength approximation). Under this approximation the homogenization procedures from the effective medium theory (is not presented here; for details see, \cite{Agranovich_SolidStateCommun_1991, Elmzughi_JPhysCondMat_1995}) are involved in order to describe the multilayered ferrite-semiconductor structure under study via an equivalent anisotropic uniform medium whose optical axis is directed along the $z$-axis. Note, the procedures are verified by the transfer matrix technique \cite{Tuz_PIERB_2012, Tuz_JOpt_2015, Tuz_Springer_2016}, and they are repeatedly used in the solid state physics \cite{Abraha_SurfSciRep_1996, Aronov_PhysRevB_1997, Kushwaha_SurfSciRep_2001, Shayesteh_PhysStatSolidi_2006, Tuz_JMMM_2016, Tuz_Superlattice_2017, Tuz_JApplPhys_2017}.

Thus, the resulting equivalent gyroelectromagnetic medium that fills the guide is further characterized by two tensors of relative effective permittivity $\hat\varepsilon_{eff}$ and relative effective permeability $\hat\mu_{eff}$ written in the form 
\begin{equation}
 \hat \varepsilon_{eff}=\left( {\begin{matrix}
   {\varepsilon} & {i\varepsilon_a} & 0 \cr
   {-i\varepsilon_a} & {\varepsilon} & 0 \cr
   0 & 0 & {\varepsilon_\|} \cr
\end{matrix}
} \right),~~~~~
\hat\mu_{eff}=\left( {\begin{matrix}
   {\mu} & {i\mu_a} & 0  \cr
   {-i\mu_a } & {\mu} & 0  \cr
   0 & 0 & {\mu_\|}  \cr
 \end{matrix}
} \right), \label{eq:eff}
\end{equation} 
where the complete expressions for the tensors components derived via underlying constitutive parameters of magnetic and semiconductor layers one can find in \cite{Tuz_JMMM_2016, Tuz_Superlattice_2017, Tuz_JApplPhys_2017}.

Thereby, the initial problem is reduced to the consideration of a circular waveguide which is completely filled by a gyroelectromagnetic uniform medium. The essence of the problem is to obtain and numerically solve the dispersion equation with a subsequent classification of the waveguide modes. The solution procedure comprises a deriving of a pair of coupled Helmholtz wave equations with respect to longitudinal components of the electromagnetic field inside the gyroelectromagnetic filling of the guide and then imposition of the boundary conditions on the PEC circular wall (the treatment is omitted here and can be found in \cite{Gurevich_book_1963, prati_JEMWA_2003}). 

\section{Classification and passbands conditions of hybrid modes}
\label{sec:res}

Our objective here is to discuss the principles of classification, determination of passbands positions and cutoff frequencies of modes of a circular gyroelectromagnetic waveguide. In order to classify the modes, it is a normal practice to consider the structure without losses (i.e. to solve the eigenwaves problem), and then determine the bands where a high level of losses can make a significant impact on the propagating waves in a practical system. 

Since in the waveguide under study the waves possess a hybrid character and arise in both symmetric and asymmetric forms, they need to be classified as hybrid HE$_{nm}$ and EH$_{nm}$ modes depending on which longitudinal component of either magnetic ($\text{H}_z > \text{E}_z$, HE-modes) or electric ($\text{E}_z > \text{H}_z$, EH-modes) field is dominant. Here the indexes $n$ and $m$ are introduced to describe the number of field variations in azimuth and radius, respectively. For the asymmetric modes ($n\ne 0$), each field variation in azimuth ($\pm n$) is related to two independent solutions of the dispersion equation, which differ by the propagation constant $\gamma$. This difference in the propagation constant distinguishes two orthogonal polarization states with the left-handed and right-handed circular rotation \cite{fuller_book_1987, Veselov_book_1988}. The handedness of the circular polarization is defined by the sign of the integer $n$. When $n<0$ the waves acquire the left-handed circular polarization (HE$_{nm}^-$ and EH$_{nm}^-$ modes), whereas when $n>0$ they are right-handed circularly polarized ones (HE$_{nm}^+$ and EH$_{nm}^+$ modes). For the symmetric modes ($n=0$), the dispersion equation has a single solution, and in this case the sign `$+$' or `$-$' in the mode notations is absent. 

The hybrid modes classification is made involving an auxiliary reference waveguide that is filled by an isotropic homogeneous medium since the modes of such a waveguide are defined definitely. Thus, for the waveguide structure under study it is supposed to start with such an isotropic (non-gyrotropic lossless) case (i.e. considering the limit $M \to 0$) and classify the modes as those of either TE-type or TM-type beginning from their cutoffs, and then gradually increase the gyrotropic parameters of the underlying constituents (i.e. non-diagonal constitutive tensors elements $\mu_a$ and $\varepsilon_a$) of the gyroelectromagnetic medium inside the guide and trace the changing in the propagation constant $\gamma$ \cite{Veselov_book_1988, whites_report_1989}. According to such scheme, hybrid modes of the HE-type and EH-type of the gyroelectromagnetic waveguide under study appear to be directly associated with corresponding modes of the TE-type and TM-type of the auxiliary reference waveguide.

In order to reveal passbands and cutoff frequencies of the hybrid modes of a gyrotropic waveguide it is convenient to introduce some generalized parameters \cite{Tuz_JEMWA_2017}, namely effective transverse permeability $\mu_\bot$, effective transverse permittivity $\varepsilon_\bot$, and effective refractive index $\eta^\pm$ written in the form: 
\begin{equation}
\mu_\bot = \mu-\mu_a^2/\mu,~~~~
\varepsilon_\bot = \varepsilon-\varepsilon_a^2/\varepsilon,~~~~ 
\eta^\pm=\sqrt{(\mu\pm\mu_a)(\varepsilon\pm\varepsilon_a)}, 
\label{eq:GeneralFunct}
\end{equation}
where in the latter term the upper sign `$+$' and lower sign `$-$' are related to the right-handed and left-handed circularly polarized waves propagating through an unbounded gyrotropic medium, respectively. 

\begin{figure}[!htbp]
\centering
\includegraphics[width=1.0\linewidth]{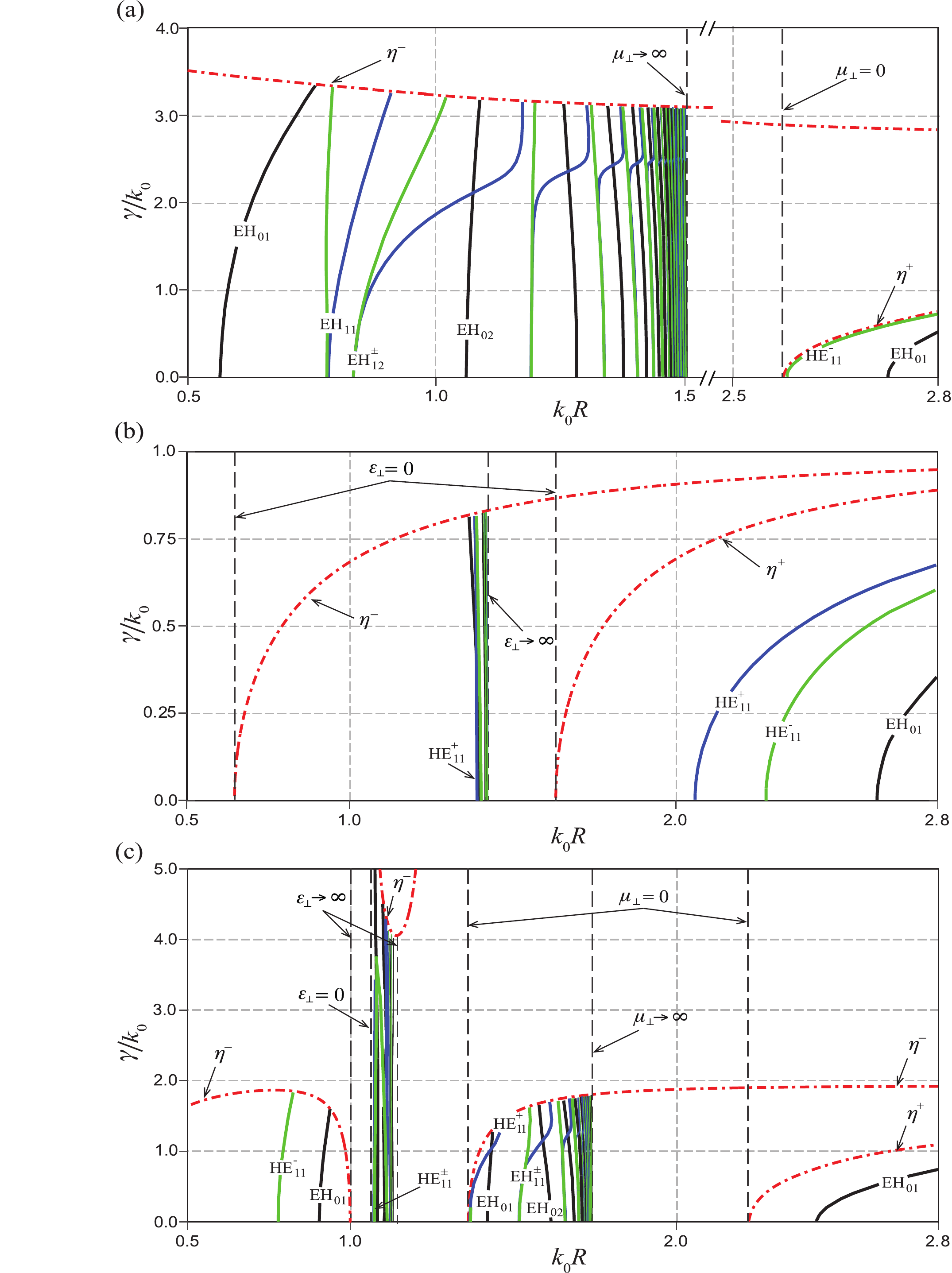}
\caption{Dispersion curves of the hybrid HE$_{nm}^\pm$ and EH$_{nm}^\pm$ modes of a circular waveguide completely filled by lossless ($b=0$, $\nu=0$) (a) gyromagnetic, (b) gyroelectric, and (c) gyroelectromagnetic medium. The blue curves and upper sign `$+$' correspond to the asymmetric modes with the right-handed circular polarization, whereas the green curves and upper sign `$-$' correspond to the asymmetric modes with the left-handed circular polarization. Black curves represent dispersion of the symmetric HE$_{0m}$ and EH$_{0m}$ modes. Extreme states $\mu_\bot=0$, $\mu_\bot \to \infty$ and $\varepsilon_\bot = 0$, $\varepsilon_\bot \to \infty$ are presented by the dashed vertical lines. All other constitutive parameters of ferrite and semiconductor are fixed as in Figure~\ref{fig:fig_1}.}
\label{fig:fig_2}
\end{figure}

The passbands of hybrid modes in a circular waveguide filled by a gyrotropic medium (i.e., it can be a gyroelectric medium described by a permittivity tensor, a gyromagnetic medium described by a permeability tensor, as well as a gyroelectromagnetic medium described by both permittivity and permeability tensors) exist inside the areas of $\gamma-k_0$ space where in the lossless case the effective refractive index $\eta^\pm$ acquires real numbers \cite{Tuz_JOpt_2010}. It corresponds to the propagation conditions of the right-handed and left-handed circularly polarized eigenwaves of a gyrotropic medium (see areas outlined by the red dash-dotted lines in Figure~\ref{fig:fig_2}). Moreover, these passbands are restricted laterally by the lines where effective transverse permeability $\mu_\bot$ or effective transverse permittivity $\varepsilon_\bot$ reaches some extrema (see vertical gray dash lines in Figure~\ref{fig:fig_2}). 

In order to highlight the main differences between dispersion characteristics of the waveguide filled by a gyroelectromagnetic medium compared to those filled by ferrite or plasma ones, the passbands of hybrid modes supported by the circular waveguide filled by a lossless gyromagnetic ($\delta_f=1$, $\delta_s=0$), gyroelectric ($\delta_f=0$, $\delta_s=1$), as well as gyroelectromagnetic ($\delta_f = \delta_s = 0.5$) medium are depicted in panels (a), (b) and (c) of Figure~\ref{fig:fig_2}, respectively. One can conclude that while in the waveguides filled by either gyromagnetic or gyroelectric medium there are two passbands, in the waveguide filled by the gyroelectromagnetic medium four passbands appear where supported modes can exist. For all types of the waveguide filling every passband starts at the frequency where $\mu_\bot = 0$ or $\varepsilon_\bot = 0$, and it ends at the frequency where the asymptotic line $\mu_\bot \to \infty$ or $\varepsilon_\bot \to \infty$ arises.

Furthermore, in each identified passband the low-index modes are properly classified assuming the blue and green curves in the figure express dispersion characteristics of the asymmetric modes with the right-handed and left-handed circular polarization, respectively, whereas the black curves represent those of the symmetric HE$_{0m}$ and EH$_{0m}$ modes. It is revealed that in the lossless case for the waveguide modes to exist both effective transverse permeability $\mu_\bot$ and effective transverse permittivity $\varepsilon_\bot$ must be positive quantities, whereas their underlying constitutive parameters $\mu$ and $\varepsilon$ can possess different signs. Remarkable, in those bands where parameters $\mu$ and $\varepsilon$ have different signs the modes exhibit an anomalous dispersion line. 

Finally, taking into account the above discussed scheme of the hybrid modes classification with involving an isotropically filled auxiliary reference waveguide, the cutoff frequency of a particular supported mode can be estimated from the following expressions \cite{Veselov_book_1988}:
\begin{equation}
\begin{split}
\text{HE-modes:}~~~~~f_{nm}^{HE}=\chi_{nm}^\prime c\left( 2 \pi R\sqrt{\varepsilon_\bot \mu_{||}} \right)^{-1}, \\ 
\text{EH-modes:}~~~~~f_{nm}^{EH}=\chi_{nm} c\left( 2 \pi R\sqrt{\varepsilon_{||} \mu_\bot} \right)^{-1},
\end{split}
\label{eq:CutOff}
\end{equation}
where $\chi_{nm}$ and $\chi_{nm}^\prime$ are zeros of the Bessel function $J_{n}$ of the first kind of order $n$ ($n=0,1,2,...$) and its first derivative $J_{n}^\prime$, respectively.  

From this estimation it is revealed that in the circular waveguide filled by a gyrotropic media an extraordinary dispersion feature may appear consisting in arising the EH$_{01}$ mode below the HE$_{11}$ mode cutoff (see, Figure~\ref{fig:fig_2}a). We also distinguish a particular mode as an isolated one if it stands alone within a certain passband of the circularly polarized eigenwaves of the unbounded gyrotropic medium.

\section{Dispersion control of the EH$_{01}$ and HE$_{11}$ modes}
\label{sec:disp}

Hereinafter it is our goal to demonstrate an ability to realize a single-mode operation for either EH$_{01}$ mode or isolated EH$_{11}$ mode in the waveguide system under study. In order to reach conditions for such an operation both the filling factors of the composite medium and radius of the guide need to be carefully adjusted to distinct cutoff of the higher order modes. In particular, compared to the results of our reference paper \cite{Tuz_JEMWA_2017}, here the waveguide radius is halved and fixed, and then an optimization problem is solved with respect to the filling factors $\delta_f$ and $\delta_s$ to find the single-mode operation conditions. 

\begin{figure}[!htbp]
\centering
\includegraphics[width=0.7\linewidth]{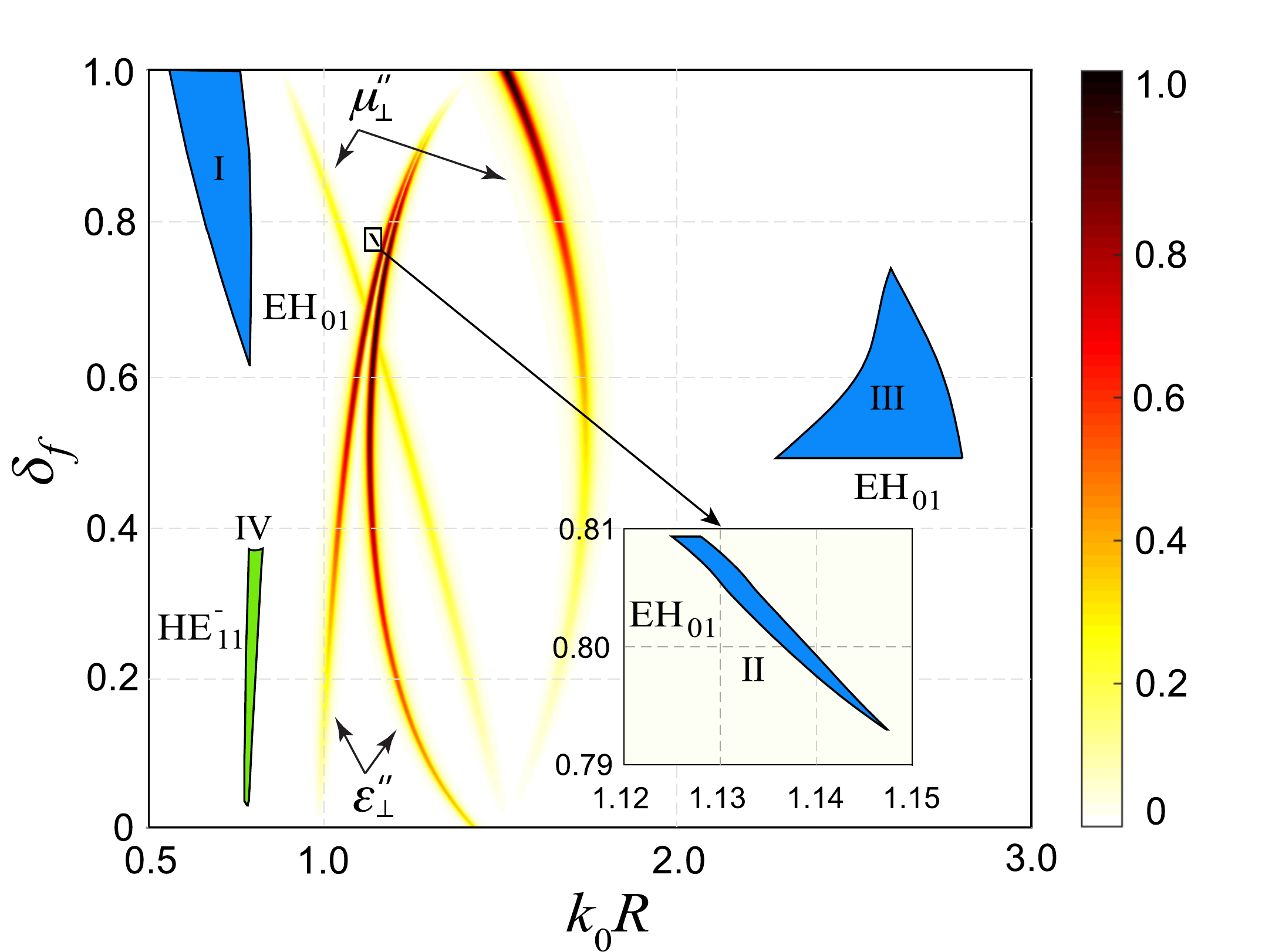}
\caption{Frequency bands versus filling factor $\delta_f$ of the gyroelectromagnetic filling of the waveguide corresponding to a single-mode operation of the EH$_{01}$ mode (blue areas) and isolated HE$_{11}^{-}$ mode (green area). Color map and intensity bar represent the imaginary parts of effective transverse permeability $\mu_\bot''$ and effective transverse permittivity $\varepsilon_\bot''$. All other constitutive parameters of ferrite and semiconductor are fixed as in Figure~\ref{fig:fig_1}.}
\label{fig:fig_3}
\end{figure}

A graphical solution of the designated optimization problem is depicted in Figure~\ref{fig:fig_3}, where the frequency bands allowable for the single-mode operation of the EH$_{01}$ mode as well as isolated HE$_{11}^{-}$ mode are identified depending on the filling factor $\delta_f$. One can conclude that in the waveguide filled by a gyroelectromagnetic medium with a predominant impact of the ferrite subsystem ($\delta_f > \delta_s$) the EH$_{01}$ mode appears to be the fundamental one within the range of filling factor $0.64\leq \delta_f \leq 1$. In particular, the conditions of the single-mode operation of the EH$_{01}$ mode are found may occur in three particular passbands. In Figure~\ref{fig:fig_3} these passbands are colored in blue and denoted by Roman numerals I, II, and III. On the other hand, the isolated HE$_{11}^{-}$ mode arises in the waveguide filled by a gyroelectromagnetic medium with a predominant impact of the semiconductor subsystem ($\delta_s > \delta_f$). Within the frequency band of interest, only one passband is found where such isolated single-mode operation conditions for the  HE$_{11}$ mode can be achieved. It appears in the range of filling factor $0.12\leq \delta_s \leq 0.43$. In Figure~\ref{fig:fig_3} this distinct passband is colored in green and denoted by Roman numeral IV.

From estimations (\ref{eq:CutOff}) it can be determined which hybrid mode (either HE$_{11}$ mode or EH$_{01}$ mode) is the fundamental one in the waveguide under study comparing values of relations $\chi_{11}^\prime/\sqrt{\varepsilon_\bot \mu_{||}}$ and  $\chi_{01}/\sqrt{\varepsilon_{||} \mu_\bot}$ obtained for the HE$_{11}$ mode and  EH$_{01}$ mode, respectively. Since $\chi_{11}^\prime$ and $\chi_{01}$ are constant quantities related to the Bessel function roots, the cutoffs of corresponding modes depend on the multipliers $\varepsilon_\bot \mu_{||}$ and $\varepsilon_{||} \mu_\bot$ in the denominators only. Thus, in the case when $\chi_{11}^\prime/\sqrt{\varepsilon_\bot \mu_{||}}  < \chi_{01}/\sqrt{\varepsilon_{||} \mu_\bot}$ the HE$_{11}$ mode is the fundamental one, otherwise it is substituted by the EH$_{01}$ mode. This remarkable feature distinguishes a gyrotropic waveguide from conventional circular hollow or isotropically filled waveguides where the EH$_{01}$ mode cannot be the fundamental one.

\begin{figure}[!htbp]
\centering
\includegraphics[width=1.0\linewidth]{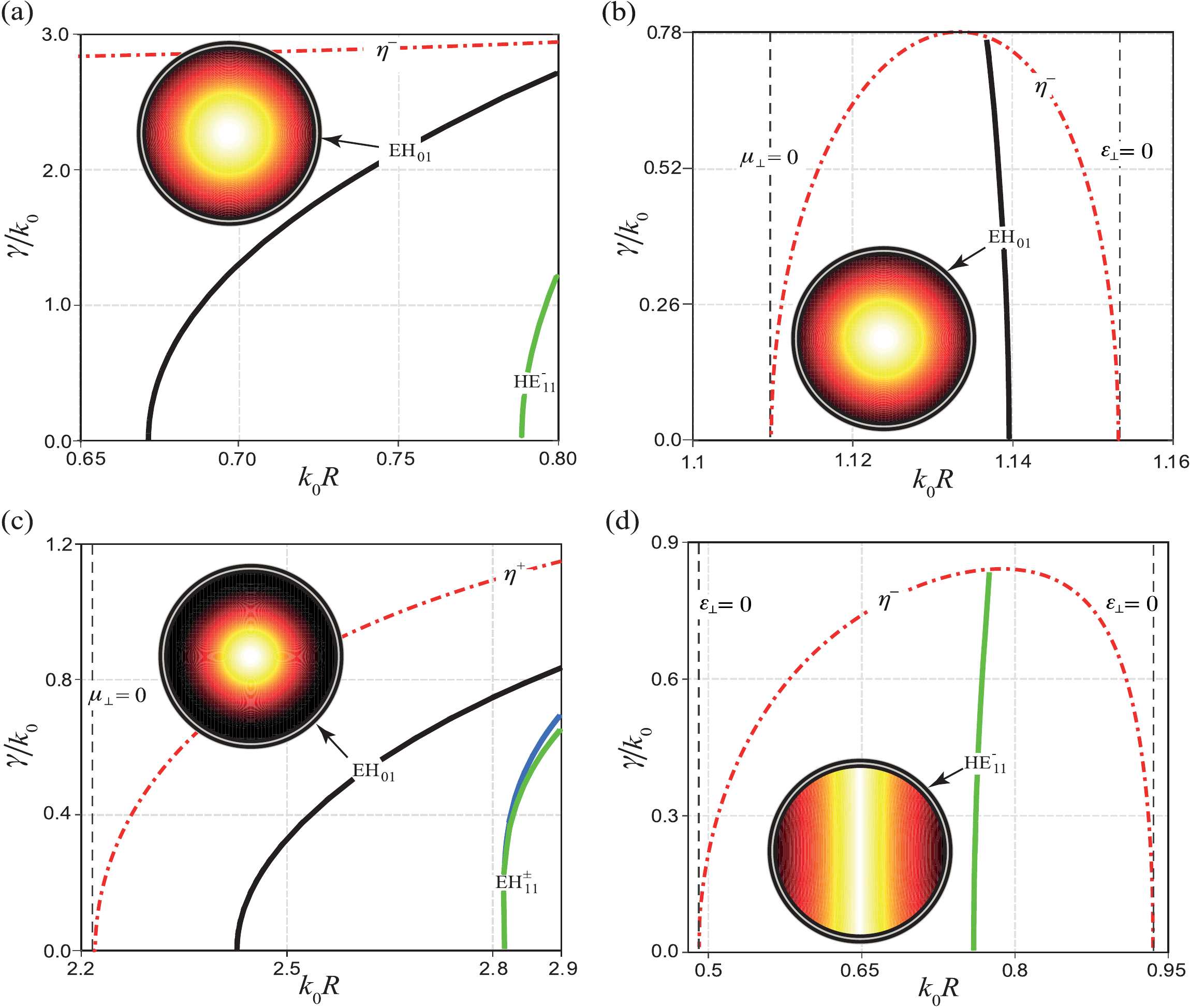}
\caption{Dispersion curves of the hybrid EH$_{01}$ and HE$_{11}^{-}$ modes, and corresponding magnitude patterns of the $z$-component of (a-c) magnetic and (d) electric field in a circular waveguide completely filled by a gyroelectromagnetic medium. The medium filling factors are: (a, b) $\delta_f = 0.8$, $\delta_s = 0.2$; (c) $\delta_f =\delta_s =0.5$; (d) $\delta_f =0.15$, $\delta_s =0.85$. All constitutive parameters of ferrite and semiconductor are fixed as in Figure~\ref{fig:fig_1}.}
\label{fig:fig_4}
\end{figure}

In order to reveal propagation conditions for the identified modes in the frequency band of interest the imaginary parts of effective transverse permeability $\mu_\bot''$ and effective transverse permittivity $\varepsilon_\bot''$ are plotted as color maps in the $\delta_f - k_0$ space where the level of losses is denoted in the color bar on the right side of Figure~\ref{fig:fig_3}. One can conclude that passbands I, III and IV lay far from the characteristic frequencies of the ferromagnetic and plasma resonances of the underlying materials, therefore, a low level of losses is expected to be in these passbands. Unfortunately, passband II is deprived of this advantage, and thus the wave propagation within this band may be significantly suppressed.

In order to verify the scheme of the hybrid modes classification, the field patterns of particular modes in the identified passbands I--IV  are plotted in Figure~\ref{fig:fig_4}. These pattern are obtained for certain values in the $\gamma-k_0$ space which are selected in the middle of the corresponding dispersion curve for different ratio of the filling factors $\delta_f$ and $\delta_s$. One can see that despite of the fact that the difference in the ratio of material fractions significantly influences the dispersion curves appearance of both hybrid EH$_{01}$ and HE$_{11}$ modes, their field patterns acquire distributions which are very similar to those of a convenient hollow or isotropically filled circular waveguide, and, thus, the hybrid modes can be identified definitely.

Moreover, from the appearance of the dispersion curves in Figure~\ref{fig:fig_4} it is revealed, that the modes can possess either normal or anomalous dispersion line depending on the characteristics of the gyroelectromagnetic filling medium. For instance, since in the particular region II the conditions $\mu>0$ and $\varepsilon<0$ hold, the isolated EH$_{01}$ mode acquires an anomalous dispersion within the narrow frequency band (see, Figure~\ref{fig:fig_4}b).

\section{Conclusion}
\label{sec:concl}
Dispersion characteristics of a circular waveguide completely filled by a longitudinally magnetized gyroelectromagnetic medium are studied. It is shown that in such a waveguide the single-mode passbands for the EH$_{01}$ and HE$_{11}$ modes can be achieved far below their natural cutoff frequencies without significant distortion of the modes field patterns. 

By solving an optimization problem, it is revealed that cutoff frequencies of both the HE$_{11}$ and EH$_{01}$ modes can be effectively controlled by adjusting the ratio between the ferrite and semiconductor components of the constitutive layers within the period of the composite medium for a fixed radius of the guide. Two remarkable results are distinguished. First, the EH$_{01}$  mode is obtained to be the fundamental one in the structure with a predominant impact of the ferrite subsystem, and, for this mode the single-mode propagation conditions accompanied by an anomalous dispersion line are achieved in the frequency band where constitutive parameters of the underlaying materials possess different signs. Second, for the fundamental HE$_{11}$ mode the single-mode operation is obtained in the structure with a predominant impact of the semiconductor subsystem.  

\section*{Disclosure statement}
No potential conflict of interest was reported by the authors.

\section*{ORCID}
\textit{Volodymyr I. Fesenko} http://orcid.org/0000-0001-9106-0858 

\bibliographystyle{tEWA}
\bibliography{GW-modes}

\noindent\medskip

\label{lastpage}

\end{document}